\documentclass[twocolumn]{article}

\usepackage{graphicx}

\begin{document}

\bibliographystyle{unsrt}

\title{Photoluminescence studies  study of a phenyl-substituted PPV}

\author{  {\small
Vladimir~A.~Sautenkov$^{\dagger\diamond}$, Steven~A. van den Berg$^{\dagger}$,
Gert~W.~'t~Hooft$^{\dagger\ddagger}$, and Eric~R. Eliel$^{\dagger}$\footnote{fax:
 +31-71-5275819, electronic address:
eliel@molphys.leidenuniv.nl}}
\\                          {\small
$\dagger$~Huygens Laboratory,
Leiden University, P.O.Box 9504, 2300 RA Leiden,
The Netherlands}
\\
{\small $\ddagger$ Philips Research Laboratories, Prof. Holstlaan 4, 5656 AA Eindhoven, The Netherlands }\\
{\small $^\diamond$ Lebedev Institute, Leninsky Prospect 53, Moscow 119991, Russia} }\date{}\maketitle
\begin{abstract}
We report on the photophysics of a phenyl-substituted PPV both in
solution and as a film. For both systems we have studied the decay of the photoluminescence
and of the emission anisotropy for a large set of wavelengths spanning the
entire photoluminescence spectrum. At long wavelengths the decay behavior
is that of an interchain species. At the shortest wavelengths the decay of the
photoluminescence from the film is observed to have a long-lived component, in
addition to the rapidly decaying component usually associated with
energy transfer.
 We attribute this slow component to emission by isolated intrachain excitons with
 reduced nonradiative relaxation.
\end{abstract}

\section{Introduction}
The first operation of a light-emitting diode based on
electroluminescence from poly($p$-phenylene-vinylene),
PPV~\cite{Burroughes1990}, has been followed by a worldwide
effort to discover new and efficient organic electroluminescent
materials, to develop practical devices, and to understand their conducting and luminescent
properties.

Notwithstanding an intense effort, the photophysics of these materials is
not yet completely understood. This is well exemplified by the recent work
on, presumably, simple systems, i.e., low-concentration solutions of MEH-PPV
\cite{Nguyen1999,Chang2000,Collison2001a}. These studies brought to light that
the solvent quality has a major influence on the spectrum, decay and quantum
efficiency of the photoluminescence. The effects are believed to be induced by
solvent-induced conformational changes of the polymer, leading to aggregate
formation in poor solvents.

In high-quality solvents, at low concentration the photoluminescence (PL) is
primarily due to the decay of the
primary exciton~\cite{Nguyen1999}, the spectrum showing a 0-0 band and
 well-resolved vibronic sidemodes. The PL quantum yield is high ($>50 \%$)  and the decay
time of the photoluminescence is of order 0.3-1.2~ns. The optical gain can
easily be made sufficiently large to generate mirrorless
lasing~\cite{Vandenberg1999}. It decays at the same rate as the photoluminescence
when the pumping strength is such that depletion of the excited state by
stimulated emission can be neglected~\cite{Vandenberg2001a}.

In the technologically relevant case that the polymer forms a thin film the
photophysics continues to be a subject of debate. For instance, the exciton
binding energy is advocated to be of order $0.1$ eV~\cite{Moses2000}, $0.5$
eV~\cite{Alvarado1998,Vanderhorst2001}, or $1$ eV~\cite{Leng1994}. This
variability makes it virtually impossible to estimate the probability of exciton
dissociation at any temperature. There is universal understanding, however, that
interchain species play an important role in the emission process of conjugated
polymer films, whether these interchain species emit themselves or not. It has
also become clear that the luminescent properties of conjugated polymer films can
be heavily influenced by the film-processing pocedure, and, again, the quality
of the solvent when the film is cast~\cite{Collison2001}.

In comparison to high-quality solutions, {\em films} of, for instance, MEH-PPV ( an alkoxy-substituted PPV)
exhibit considerably reduced PL efficiency, a PL decay which is slower and nonexponential, and very rapid optical
gain dynamics. Although these characteristics are found for many PPV-derivatives, they are not universal. For
instance, in DP6-PPV, the gain dynamics of the film is as slow as that of the solution~\cite{Dogariu2000}, making
this material well-suited as gain material in a conjugated-polymer-film laser. Also, a substantial number of
PPV-derivatives, notably the phenyl-substituted varieties, have roughly equal, high, PL quantum-efficiencies in
the film and in solution. These phenyl-substituted PPVs deserve particular attention, being the first to be used
in a commercial product that is on the market at the present time.

Notwithstanding their market introduction, the photophysics of such phenyl-substituted
PPVs, developed to have a high electroluminescent efficiency in the condensed
phase~\cite{Spreitzer1998}, has hardly been studied. To address that issue we
report here results on PL spectra,  spectrally resolved PL
decay, and the time and spectrally resolved PL anisotropy for both
a neat film and a solution of the polymer.

\section{Experiment}
The conjugated polymer under study is the phenyl-substituted PPV,
poly[2-phenyl-(2'-decyloxy)-1,4-phenylene-vinylene]. It exhibits
good solubility in standard organic solvents and has a high
electroluminescent efficiency when made into a thin
film~\cite{Spreitzer1998}. Our base material is a solution of the
polymer in chlorobenzene at a concentration of 1 g/l. We fill a
small cuvette with the solution to measure the PL spectrum and PL
decay of the solution. Additionally, we form thin films by spin
coating  the solution  on a BK7 substrate in a dry nitrogen
atmosphere. The samples are kept and studied permanently under
such atmosphere to avoid degradation due to the presence of oxygen
and/or water. The optical density of the films is measured to be
approximately 2 at a wavelength of 400 nm.
\begin{figure}[t]
\centerline{\includegraphics[width=7cm]{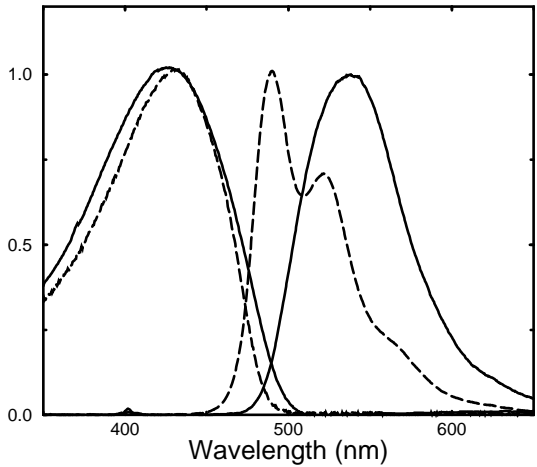}} \caption[]{Absorption and photoluminescence spectra of
the phenyl-substituted PPV polymer under study. The dashed lines show the spectra for the polymer dissolved in
chlorobenzene while the solid lines display the absorption and emission spectrum of a spin-coated polymer film.}
\label{ch.anisotropy;Fig.spectra}
\end{figure}

We excite the polymer, in solution or as a thin film, with
the frequency-doubled output ($\lambda=400$ nm) of an ultrafast
(60 fs pulse duration) Ti:sapphire laser (Kapteyn-Murnane
Laboratories) operating at a repetition rate
 of 82~MHz. The excitation pulses from the laser are heavily attenuated in order
 to remain in the linear-response regime and to avoid degradation of the sample.
 The laser beam is vertically polarized and is focussed to a spot of 1~mm
 diameter at the sample. The photoluminescence emitted by the polymer is collected
  with a lens and imaged on the input aperture of our detection equipment.
 Steady-state PL spectra are recorded with a simple fiber-optic spectrometer
 (Ocean Optics S2000), while we use time-correlated single-photon counting to measure
  PL in the time domain. With a multichannel-plate based photomultiplier
  (Hamamatsu model R3809U-52) our setup has a temporal resolution of $\approx 40$ ps. In the
  time-domain measurements we use spectral filters (bandwidth 10 nm, $10^{-7}$ out-of-band
  transmission) to
  select  specific wavelength bands in the range 472-600 nm.
  We include a polarizer in our setup to separately record
  the polarized and sensitized luminescence.

Figure~\ref{ch.anisotropy;Fig.spectra} shows the steady-state
absorption spectrum together with the time-integrated PL spectrum
of the polymer, both in solution (dashed curves) and as a thin
film (solid lines).
 For the solution the absorption spectrum peaks at $\lambda=430$ nm,
 whereas the PL spectrum has its maximum at $\lambda=490$ nm. The
absorption spectrum of the film is similar to that of the
solution, peaking at
 $\lambda=428$ nm. As compared to that of the solution, the PL spectrum of the
 film is red shifted, relatively broad and structureless, with a peak at
 $\lambda=538$ nm. When excited with more powerful pump pulses, generated by an
 amplified laser system, we observe
  amplified stimulated emission from our film, centered around  $\lambda=545$ nm.

  \section{Photoluminescence decay}
\begin{figure}[t]
\centerline{\includegraphics[width=7cm]{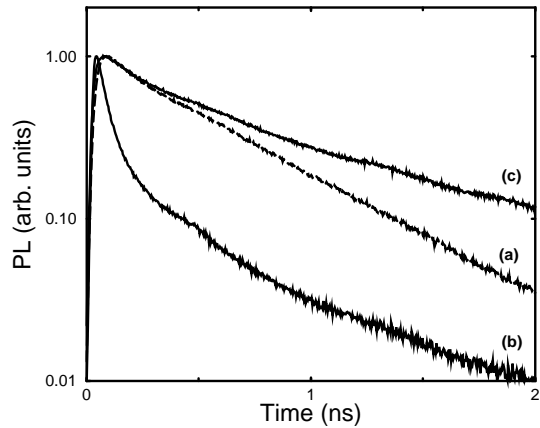}} \caption[]{Decay of the photoluminescence of the polymer
as excited by an ultrashort pulse at $\lambda= 400$ nm. Curve (a) shows the emission at $\lambda=472$ nm by the
dissolved polymer while curves (b) and (c) give the results for the polymer film at a wavelength of $\lambda=472$
nm and $\lambda=600$ nm, respectively.
  }
\label{ch.anisotropy;Fig.pldecay}
\end{figure}

For the polymer {\em in solution} the temporal evolution of the PL
is almost wavelength independent and is well described by a single
exponential:
\begin{equation}
S=S_0 \exp(-\gamma t),\label{ch.anisotropy;Eq1}
\end{equation}
where $\gamma$ represents the PL decay rate. The decay takes place
due to a combination of radiative (spontaneous emission) and
nonradiative processes, which have, for the polymer at hand,
roughly equal rates. Curve (a) in
Fig.~\ref{ch.anisotropy;Fig.pldecay} shows the decay of the PL of
the dissolved polymer as measured in a narrow interval around
$\lambda=472$ nm, i.e., in the blue wing of the pure
singlet-exciton transition (see
Fig.~\ref{ch.anisotropy;Fig.spectra}).

The PL decay of the {\em thin film} shows a strong wavelength
dependence, and can {\em not} be described by a single
exponential. Rather, the PL decay shows both a fast and slow
component, the latter  becoming slower with increasing wavelength.
Additionally, for the longer emission wavelength, one observes
that the PL rise time is no longer instrumentally limited as is
the case for shorter emission wavelengths.
 To illustrate the wavelength dependence of the PL decay we include in
 Fig.~\ref{ch.anisotropy;Fig.pldecay} decay curves
 for the thin film (solid lines), as recorded in wavelength bands around $\lambda=472$
 (curve(b))and $\lambda=600$ nm (curve(c)). To simplify the comparison of the
 various decay curves in Fig.~\ref{ch.anisotropy;Fig.pldecay}, they have all been normalized to unity at
 their respective maxima.

 In the standard description of light emission by conjugated polymers {\em intrachain}
 energy
migration and relaxation play an important
role~\cite{Kersting1993}. The singlet exciton, generated through
the photo-excitation process, is thought to migrate from regions
of short conjugation length to regions of longer conjugation
length, thereby reducing its energy. This process takes place on a
time scale of order $100$ ps~\cite{Schwartz2001}. Additionally, there can be {\em
interchain} energy transfer due to dipole-dipole coupling, for
instance through the F\"{o}rster process~\cite{Forster1948}. This
model is  generally believed to be incomplete. Experimental
evidence abounds that one has to assume the presence of at least
two types of excitation: intrachain excitons and long-lived,
red-emitting interchain species, such as excimers. These
excitations
 are coupled through a one-way transfer from exciton to excimer. This transfer
 takes place in a time comparable
 to the response time of our setup ($\approx 40$ ps). Note that not all primary
 excitons need to follow this path. The presence of a long-lived component in
 the film PL at $\lambda=472$ nm suggests that at least some of the primary excitons survive.

 We  parametrize the PL by a biexponential:
\begin{equation}
S(t,\lambda)= S_{\rm fast}^\lambda\exp (-\gamma_{\rm fast}^\lambda
t) + S_{\rm slow} ^\lambda\exp(-\gamma_{\rm slow}^\lambda
t),\label{ch.anisotropy;Eq3}
\end{equation}
where $ S_{\rm fast}^\lambda$ and $S_{\rm slow}^\lambda$ are
proportional to the population of the fast-decaying and the
slow-decaying species that emit at wavelength $\lambda$,
respectively. These species decay at rates $\gamma_{\rm
fast}^\lambda$  and $\gamma_{\rm slow}^\lambda$ respectively.
Figure~\ref{ch.anisotropy;Fig.pldecay} shows that the slowly
decaying component can indeed be described by an exponential. Our
parametric description of the luminescence fits our experimental
data quite well, particularly at longer wavelengths.

For $\lambda>515$ nm we associate the slowly decaying component of
the PL with the excimer, and the component with fast decay with
the exciton, each emitting at that specific wavelength. In the
extreme blue wing of the PL spectrum we make a different
connection; there we associate the fast decaying component of the
PL with the (nonradiative) relaxation of the bulk of the exciton
population, while the slow decay represents the evolution of a
class of high-energy excitons that have strongly reduced
nonradiative decay; henceforth we will refer to these excited
states as ``isolated excitons''. The rapid decay of the bulk of
the population of high-energy excitons is due to nonradiative
processes such as intrachain and interchain energy transfer and
coupling to excimer states. In that light it does not come as a
surprise that Eq.~(\ref{ch.anisotropy;Eq3}) does not fit our data
so well at short wavelengths. One can hardly expect the
complicated energy transfer process to be described by a bi-exponential.
\begin{figure}[ht]
\centerline{\includegraphics[width=7cm]{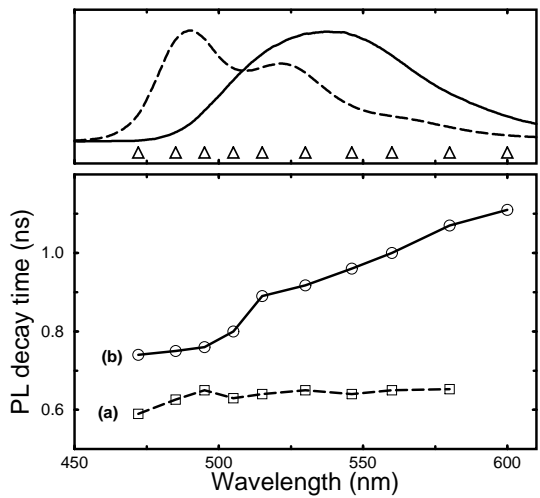}}
 \caption[]{Upper frame: PL spectra measured  for a
solution (dashed line) and for a film (solid line). The triangles indicate the midband positions of the
narrow-bandwidth (10 nm) filters to selectively measure the wavelength dependent PL decay. Lower frame: The dashed
curve shows the wavelength dependence of the decay time of the PL from a solution; the solid curve displays the
behavior of the decay constant associated with the long tail of the PL emitted by a film.}
\label{ch.anisotropy;Fig.decaytimes}
\end{figure}

From the fits one can extract values for the two decay rates
  $\gamma_{\rm fast}^\lambda$  ($\le 5\times 10^9$ s$^{-1}$)
and  $\gamma_{\rm slow}^\lambda$. In
Fig.~\ref{ch.anisotropy;Fig.decaytimes} we show the values of
$\tau_{\rm slow}(\lambda)=(\gamma_{\rm slow} ^\lambda)^{-1}$ as a
function of emission wavelength.  For reference purposes we have
included the decay time  for the polymer in solution as obtained
from a fit of the appropriate data with a curve described by
Eq.~(\ref{ch.anisotropy;Eq1}).
        Figure~\ref{ch.anisotropy;Fig.decaytimes} also shows that the film
PL (at least its slow component) decays at a rate that is
appreciably slower than that of the solution. This observation
supports our assumption that an interchain species is involved
in the slow decay of the film PL.

\section{Photoluminescence anisotropy}
Additional information about interchain interactions can be
obtained from a time-domain study of the photoluminescence
anisotropy~\cite{Herz2001,Watanabe1997}, given by:
\begin{equation}
r(t,\lambda)  = \frac{S_\parallel (t,\lambda) - S_\perp
(t,\lambda)} {S_\parallel (t,\lambda) + 2S_\perp
(t,\lambda)}.\label{ch.anisotropy;Eq4}
\end{equation}
Here $S_\parallel(t,\lambda)$ and $S_\perp(t,\lambda)$ represent
the strength of the polarized and sensitized PL signals at
wavelength $\lambda$ as a function of time.
 \begin{figure}[ht]
\centerline{\includegraphics[width=7cm]{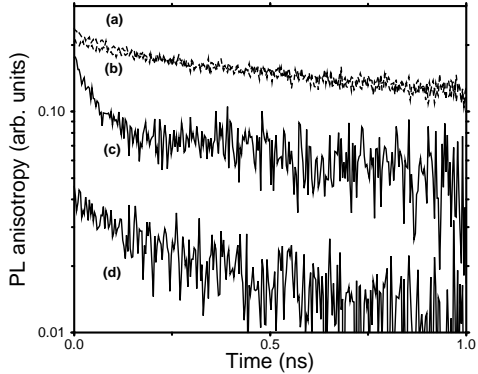}}
 \caption[]{ Time-dependent anisotropy of the emission
from a solution (curve (a) for $\lambda=472$ nm and curve (b) for $\lambda=580$ nm), and from a film (curve (c)
for $\lambda=472$ nm and curve (d) for $\lambda=600$ nm). } \label{ch.anisotropy;Fig.anisotropydecay}
\end{figure}

Modeling the conjugated polymer, both in solution and in the film, as an ensemble
 of randomly oriented dipoles one finds
$r(0)= 0.4\ $(that is $S_\parallel(0)/S_\perp(0) = 3)$, if the absorbing and
emitting dipoles are parallel. Beyond $t=0$, the photoluminescence
anisotropy (PLA) decays due to orientational relaxation. In a
polymer solution this happens through orientational diffusion of
the polymer chains~\cite{VandenBerg2001b}. In the film, the decay
of $r(t,\lambda)$ is due to intrachain and interchain excitation
transfer, or creation of interchain states.

In our experiments we find that $r(t,\lambda)$ never exceeds the
value $0.3$, even for the solution at $t=0$. This implies either
misalignment of the absorbing and emitting dipoles, or very fast
depolarization during the energy relaxation process from the photo-excited
to the luminescent level of the singlet manifold.

Because of the importance of the instrumental response during the
rising part of the PL signal, one can not arrive at a reliable
value of $r(t, \lambda)$
 while either $S_\parallel(t,\lambda)$ or $S_\perp(t,\lambda)$ increase rapidly.
 We therefore  calculate $r(t,\lambda)$ only during the decay of the PL signals,
 with $t=0$ being defined as the time at which the PL signals reach their maximum.
Figure~\ref{ch.anisotropy;Fig.anisotropydecay} shows our results
for the time-dependent anisotropy for the same set of wavelengths as in
Fig.~\ref{ch.anisotropy;Fig.pldecay}.
                       In all measurements the PLA
decays as a function of time, not reaching a steady-state value
within our time window (1~ns). Additionally, the PLA of the
solution is, at all times, substantially larger than that of the
film. For $t\ge 0.2$~ns the  PLA is well described by an
exponential time dependence; this applies to all wavelengths both
in solution and in the film   By fitting our data for $t\ge
0.2$~ns with an exponential we obtain the associated decay times,
of order $1-3$~ns.
\begin{figure}[ht]
\centerline{\includegraphics[width=7cm]{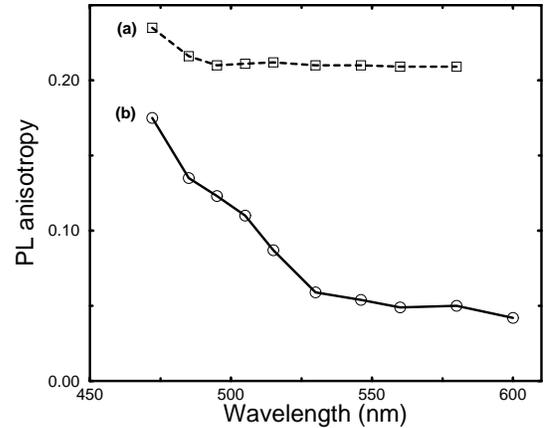}}
 \caption[]{ Wavelength dependence of the PL anisotropy
at $t=0$ for the polymer solution (dashed curve), and for a polymer film (solid curve).
}\label{ch.anisotropy;Fig.anisotropyt0}
\end{figure}

Figure~\ref{ch.anisotropy;Fig.anisotropyt0} shows the values of
$r(0)$, i.e., the maximum value of the  photoluminescence
anisotropy.
For the solution we find a, basically, wavelength independent
value of $r(0)$, except at the shortest wavelengths.  Curve (a) of
Fig.~\ref{ch.anisotropy;Fig.anisotropydecay} shows that the
increased value of $r(0)$ at $\lambda=472$ nm is  due to an
enhanced contribution of the fast-decaying component of $r(t)$.
The value of $r(0)$ for the film is strongly wavelength dependent.

\section{Discussion}
Reviewing all our experimental results for the {\em solution} we
note that the photophysics in that system is essentially {\em
wavelength independent} except at the shortest wavelengths. This
is summarized succinctly by the relevant curves of
Figs.~\ref{ch.anisotropy;Fig.decaytimes}
and~\ref{ch.anisotropy;Fig.anisotropyt0}. The photoluminescence
decay can be described by a single exponential, yielding a PL
decay time of $\approx 0.65$~ns. The photoluminescence anisotropy
is substantial and has a long-lived component that is well
described by an exponential time dependence, with a time constant
of a few~ns. This component of the PLA decay can be attributed to
orientational diffusion of the polymer in its
solvent~\cite{VandenBerg2001b}. The weak, fast decaying component
of the anisotropy is associated with
dipole-dipole-coupling-induced loss of orientation during
(intrachain) energy migration along the polymer
backbone~\cite{Nguyen1999,Watanabe1997,Hayes1997}. This is
supported by our data that show that this fast-decaying component
is quite a bit larger at the shortest wavelength than at longer
wavelengths (see Fig.~\ref{ch.anisotropy;Fig.anisotropydecay}).
The photoluminescence at short wavelengths is entirely due to
excitons that have not migrated, whereas at longer wavelengths the
emission is due to excitons that have been populated directly or
through an intrachain energy transfer process. Interchain energy
transfer can be excluded because of the low probability of
interchain overlap in solution.

The results for the {\em film} are in sharp contrast with those of
the solution. Here we observe a {\em strong wavelength dependence}
(see
Figs.~\ref{ch.anisotropy;Fig.pldecay},~\ref{ch.anisotropy;Fig.decaytimes},~\ref{ch.anisotropy;Fig.anisotropydecay}
and~\ref{ch.anisotropy;Fig.anisotropyt0}) and decay curves that
can not be described by a single exponential. As discussed earlier
this behavior is due to interchain processes, all-important in the
film, and virtually absent in the solution.

Combining the results of Fig.~\ref{ch.anisotropy;Fig.anisotropyt0}
with those of    Fig.~\ref{ch.anisotropy;Fig.spectra},
 one notices that the anisotropy is very weak
 ($r(0)<0.06$, $S_\parallel/S_\perp<1.2$) and essentially featureless
 over the main body of the emission spectrum of the film. This small value of the anisotropy
 supports the earlier made assumption that, in this wavelength interval, the emission
originates predominantly in species that have not been populated
directly in the photo-excitation process. In the extreme red wing,
where the photoluminescence decays slowest (see
Fig.~\ref{ch.anisotropy;Fig.decaytimes}), these have interchain
character and are collectively called excimers. These interchain
species are generally believed to have reduced oscillator
strengths, hence the slow decay of the photoluminescence (see
Fig.~\ref{ch.anisotropy;Fig.decaytimes}). Less far in the red wing
the emission may well be due to both excitons and excimers, the
excitonic emitting states (partly) having been populated through an
intrachain energy transfer process.

 In the extreme blue wing of the emission spectrum the PLA is strongly
 wavelength dependent, as shown by Fig.~\ref{ch.anisotropy;Fig.anisotropyt0}. Here the value of $r(0)$
 is comparable to that in solution (see Fig.~\ref{ch.anisotropy;Fig.anisotropydecay}).
 This suggests that, at these wavelengths, the same species (a singlet exciton)
 is responsible for the emission by both the film and the solution.
 Nonradiative processes such as intrachain transfer rapidly deplete
 this exciton population, inducing the fast initial decay of the PL.
 The dipole-dipole coupling that is believed to be responsible for (part of)
 these transfer process causes the observed rapid loss of anisotropy at these
 wavelengths. The slowly decaying tail of the luminescence attests to the
 presence of excited species that emit at these wavelengths at much later times.
 We believe that these are not excimers or other low-energy
 interchain species. Rather, we attribute this emission to the isolated excitons,
  introduced earlier.
 These communicate only weakly with their environment, as shown by their slow
 nonradiative decay and slow depolarization.

It is interesting to make a connection to measurements of the
transient gain in thin films of this and similar PPVs. Initially, the
transient gain can be quite considerable but is, almost
universally, seen to decay very rapidly, on a time scale of tens
of picoseconds. It is tempting to connect the fast decay of the
gain in polymer films with the rapid decay of the initially
excited species~\cite{Collison2001}. Hence, we postulate that the gain is exclusively
related to the exciton population, while a considerable part of
the photoluminescence is  related to the population of excimers. The
mere fact that the photoluminescence quantum efficiency of a
conjugated polymer is particularly high thus does not guarantee that it
is especially suitable as a gain material in a laser. Rather, in
order to make a good optically pumped polymer laser, one has to
reduce the interchain coupling in the film.
                             This discussion puts into context the
   recent extensive efforts to control and reduce interchain effects in
   conjugated
polymers~\cite{Nguyen2000,Whitelegg2001,Schwartz2001,Chiavarone1999,Sun2000a,Sun2000b,Jakubiak2001}.
Note, however, that interchain
   coupling is essential for any kind of charge transport, i.e., for the
   development of injection-type polymer lasers.

It is clear that a description of the lasing process in thin films
of conjugated polymers in terms of a standard four-level system
is, from this perspective,  not to the point. An interesting
issue to pursue is the question why the excimers in conjugated
polymers do not give rise to lasing.  A curious aspect of excimer
lasing in conjugated polymers is that the energy transfer process
that populates the excimer states enhances the Stokes shift.
 In standard laser physics, a large Stokes shift is considered to be advantageous.
 The reduced oscillator strength of the excimers should not pose a problem in view of the many
laser systems that are based on very low oscillator-strengths
optical transitions. Obviously, the crucial question concerns
whether the excimer emission is self-absorbed by the material,
i.e., whether (photo-induced) absorption by excimers or other
interchain species will overwhelm the gain.

\section{Conclusions}
 We have studied the photophysics of a
phenyl-substituted PPV, dissolved in a common organic solvent, and
as a thin film by performing time-domain measurements of the
spectrally resolved photoluminescence and its anisotropy across
the full emission spectrum of the material. As for many conjugated
polymers, the emission by the film is red-shifted as compared to
that of the solution. The overall picture is that the measured
properties for the dissolved polymer are wavelength independent,
while the luminescent properties of the film vary considerably
with wavelength. These observations thereby confirm the results
obtained by others in that the photoluminescence in the film has,
for the most part, a different origin as that in solution. While
in solution the emission is dominated by the radiative decay of
singlet excitons, in the film the main contribution comes from
interchain species such as excimers.

 We have paid particular attention
to the blue wing of the photoluminescence spectrum where the
differences between film and solution are smallest. In that
spectral region the film photoluminescence (and its anisotropy) is
dominated by a quickly decaying component, the decay being caused
by intrachain relaxation. Once this component has decayed away, we
measure an emission that slowly gets weaker. We attribute this
emission to isolated excitons that only weakly interact with their
surroundings.

Finally, we speculate on the connection
between photoluminescence and gain in films of conjugated
polymers. We discuss whether the long-living emitter in the film
can be exploited for lasing.

\section{Acknowledgements}
We thank Dr. H.F.M. Schoo for supplying the polymer material. This work is part
of the research programme of the Stichting voor Fundamenteel Onderzoek der
Materie (FOM, financially supported by the Nederlandse Organisatie voor
Wetenschappelijk Onderzoek (NWO))and Philips Research.

\end{document}